\documentclass[aps,prd,superscriptaddress,nofootinbib,
twoside,notitlepage]{revtex4-1}     
\usepackage{graphicx}
\usepackage{euscript,amssymb}
\usepackage{amsfonts}
\usepackage{amssymb}
\usepackage{amsmath}
\usepackage{color}
\usepackage{epstopdf} 
\usepackage{subcaption}

\usepackage{amsthm}
\theoremstyle{criterion}
\newtheorem*{criterion1}{Criterion 1}
\newtheorem*{criterion2}{Criterion 2}
\newtheorem*{criterion3}{Criterion 3}
\newcommand{\ben}{\begin{eqnarray*}}
 \newcommand{\een}{\end{eqnarray*}}
\newcommand{\bdm}{\begin{displaymath}}
\newcommand{\edm}{\end{displaymath}}
\newcommand{\dd}{\textnormal{d}}
\newcommand{\ee}{\textnormal{e}}
\newcommand{\ii}{\textnormal{i}}
\newcommand{\ww}{\textnormal{w}}
\newcommand{\dif}[3]{\frac{\partial^{#3} #1}{\partial #2^{#3}}}

\DeclareMathOperator\arcsinh{arcsinh}

\begin{document}

\title{Singularity avoidance in Bianchi I quantum cosmology}

\author{Claus Kiefer}

\email{kiefer@thp.uni-koeln.de}

\author{Nick Kwidzinski}

\email{nk@thp.uni-koeln.de}

\author{Dennis Piontek\footnote{Present address: Deutsches Zentrum
    f\"ur Luft- und Raumfahrt, Institut f\"ur Physik der Atmosph\"are,
    Oberpfaffenhofen, Germany}} 

\email{dennis.piontek@dlr.de
} 

\affiliation{Institute for Theoretical Physics, University of Cologne,
Z\"{u}lpicher Stra\ss e 77, 50937 K\"{o}ln, Germany}

\date{\today}

\begin{abstract}
We extend recent discussions of singularity avoidance in quantum
gravity from isotropic to anisotropic cosmological models.
The investigation is done in the framework of quantum geometrodynamics
(Wheeler-DeWitt equation). 
We formulate criteria of singularity avoidance for general Bianchi
class A models and give explicit and detailed results for Bianchi~I
models with and without matter. The singularities in these cases
  are big bang and big rip.
We find that the classical
singularities can generally be avoided in these models. 
\end{abstract}

\maketitle

%%%%%%%%%%%%%%%%%%%%%%%%%%%%%%%%%%%%%%%%%%%%%%%%%%%%%%%%%%%%%%%%%%

\section{Introduction}\label{Intro}

A major issue in any quantum theory of gravity is the fate of the
classical singularities. So far, such a theory is not available in
final form, although various approaches exist in which this question
can be sensibly addressed \cite{OUP,Calcagni}. It is clear that such
an investigation 
 cannot yet be  done at the level of mathematical rigor
comparable to the singularity theorems in the classical theory (see e.g.
\cite{handbook}). Nevertheless, focusing on concrete approaches and
concrete models, one can state criteria of singularity avoidance and
discuss their implementation. This is what we shall do here.

We restrict our analysis of singularity avoidance to
quantum geometrodynamics, with the Wheeler-DeWitt equation as its
central equation \cite{OUP}. Although this may not be the most
fundamental level of quantum gravity, it is sufficient for addressing the
issue of singularity avoidance. Quantum geometrodynamics follows
directly from general relativity by rewriting the Einstein equations
in Hamilton-Jacobi form and 
formulating  quantum equations that yield the Hamilton-Jacobi
equations in the semiclassical (WKB) limit. It thus makes as much sense to
addressing singularity avoidance here than it does to 
addressing it in quantum
mechanics at the level of the Schr\"odinger equation.
Singularity avoidance has also been discussed in loop
quantum gravity \cite{Calcagni, Bojowald, Wilson-Ewing}, with various results, but
we will not consider this here.

Singularity avoidance was already addressed by DeWitt in his
pioneering paper on canonical quantum gravity \cite{DeWitt}. He
suggested to impose  the condition $\Psi\to 0$ for the
quantum-gravitational wave functional $\Psi$ 
when approaching the region of a
classical singularity. The wave functional is effectively defined on
the configuration space of all {\em three}-dimensional geometries,
also called superspace \cite{OUP,Wheeler68}. The ``DeWitt criterion''
of vanishing wave function then means that $\Psi$ must approach zero
when approaching a singular three-geometry (which itself is not part
of superspace, but can be envisaged as its boundary). It is important
to emphasize that this criterion is a sufficient but not a necessary
one: singularities can be avoided for non-vanishing or even diverging $\Psi$
(recall the ground state solution of the Dirac equation for the
hydrogen atom, which diverges). 

DeWitt had in mind cosmological singularities such as big bang or big
crunch. The DeWitt criterion applies, of course, also
to the singularities 
that classically arise from gravitational collapse. In simple models of
quantum geometrodynamics, their avoidance can be rigorously
addressed. One example is the collapse of a null dust shell, which
classically develops into a black hole, but quantum gravitationally
evolves into a re-expanding shell, with $\Psi=0$ in the region of the
classical singularity \cite{HK01}. In general, however, such
cases are too difficult to allow for an exact mathematical treatment, so
most investigations so far 
were restricted to Friedmann-Lema\^{\i}tre-Robertson-Walker (FLRW)
cosmology. The first detailed discussion of singularity avoidance in
Wheeler-DeWitt quantum cosmology was performed for the big-rip
singularity that occurs in the presence of phantom matter
\cite{DKS}.\footnote{See \cite{BKM19} for a recent review of the fate
  of singularities in models with phantom matter.}
Other applications followed, which generally focused on
singularities occurring in dark-energy models but also on the big bang;
see, for example, \cite{Kam-CQG,nick,Bouhmadi15,Bouhmadi16} and the
references therein. The question was also investigated for $f(R)$
quantum cosmology \cite{ABM18}. In most cases, the DeWitt criterion
was applied. In our paper, too, the main focus will be on this
  criterion, although we shall employ a second criterion that
  makes use of a current density. 

In the present paper, we make a step forward and discuss the issue of
singularity avoidance for {\em anisotropic} cosmologies. The simplest
case is the Bianchi~I model (see e.g. \cite{Ryan_Shepley}), to which
we restrict our investigation here. 
The more interesting Bianchi~IX
model is reserved for a future investigation. Anisotropic models are
characterized by the fact that the dimension of their configuration
space (minisuperspace) is bigger than two already for the pure
gravitational case. This will be important for the formulation of the
DeWitt criterion. We address the anisotropic case here mainly
  for structural reasons, in order to see how criteria of singularity
  avoidance apply there. We do not expect anisotropies to play a
  crucial role in the late universe, although such anisotropies may be
  relevant in the early universe.

Our article is organized as follows.
In Sec.~II, we formulate our criteria for singularity avoidance. In
this, a generalization is made that takes into account the conformal
structure of minisuperspace. Section~III then addresses the vacuum
Bianchi~I (Kasner) model. There, we encounter only the big bang
  singularity. Sections IV and V are devoted to Bianchi~I
models with matter: an effective matter potential is used in Sec.~IV,
and a dynamical (phantom) scalar field is used in Sec.~V.
While in Sec.~IV both the big bang and big rip singularities are
  addressed, Sec.~V focuses on the big rip singularity. We shall
find that singularities can be avoided in all relevant
cases. Section VI presents a short conclusion and an outlook.

%%%%%%%%%%%%%%%%%%%%%%%%%%%%%%%%%%%%%%%%%%%%%%%%%%%%%%%%%%%%%%%%%%%%%%%%%%%%%%%%

\section{Criteria for singularity avoidance}\label{classical} 

In this section, we formulate the criteria of singularity avoidance at
the level of a general (diagonal) Bianchi class~A model. 
These will be applied in detail to the Bianchi~I model in the
following sections. 

The action for such Bianchi models can be brought into the form
\cite{Ryan_Shepley,Jantzen,magic} 
\begin{equation}
S_{\text{EH}}+S_{\text{m}}=\int \dd t \ N \ee^{3\alpha} \left[ 
\frac{-\dot{\alpha}^2 +\dot{\beta}_+^2 +\dot{\beta}_-^2}{2N^2} +
\frac{{}^{(3)}\!R}{12} \right]+S_{\text{m}}.
\end{equation}
We parametrize the minisuperspace $\mathcal{M}$ of these
models by the
coordinates $\mathbf{q}=\left\{\alpha , \beta_+ , \beta_-,\phi
\right\}$, where $\alpha\equiv\ln a$, $\beta_+$, and $\beta_-$ are the 
Misner variables,\footnote{But note that Misner in \cite{magic} uses
  $\Omega=-\alpha$.} and $\phi$ denotes matter field
degrees of freedom; ${}^{(3)}\!R$ is the three-dimensional Ricci
scalar. 
Units are chosen such that $3c^6V_0/4\pi
G=1$, where $V_0$ is the volume of three-dimensional space (assumed to
be compact here).

Variation with respect to the lapse $N$ yields the Hamiltonian
constraint,
\begin{equation}
  \label{constraint}
\mathcal{H}=\frac{1}{2}\mathcal{G}^{IJ}p_I p_J + \mathcal{V}=0,
\end{equation}
where the $p_I$ are the momenta canonically conjugate to the
configuration variables $\mathbf{q}$,
the $\mathcal{G}_{IJ}$ denote the components of the DeWitt metric,
$\mathcal{G}^{IJ}$ the components of its inverse, and
$\mathcal{V}$ is the minisuperspace potential which
contains contributions from the three-curvature and from the matter part. 
We remark that the equations of motion can be formulated in
configuration space as a geodesic
equation plus a forcing term (\cite{magic}, p.~452). 

Because of the constraint nature \eqref{constraint} of the
Hamiltonian, minisuperspace possesses a natural conformal
structure. This can be seen as follows.
Let us consider a rescaling of the lapse,
$N\rightarrow\widetilde{N}=\Omega^2 N$, with a differentiable function
$\Omega:\mathcal{M}\rightarrow \mathbb{R}_+$.  
The transformation of the Hamiltonian constraint then follows from the
invariance of the total Hamiltonian $H$ according to
\begin{equation}
\begin{aligned} 
H &=N\mathcal{H}=\widetilde{N}\widetilde{\mathcal{H}}
=\widetilde{N}\left(\frac{1}{2}\Omega^{-2}\mathcal{G}^{IJ}p_I p_J +
  \Omega^{-2}\mathcal{V}\right) \\ 
&
=:
\widetilde{N}\left(\frac{1}{2}\widetilde{\mathcal{G}}^{IJ}p_I
  p_J+ \widetilde{\mathcal{V}}\right).
\end{aligned} 
\end{equation}
This rescaling induces a local Weyl transformation of the DeWitt metric,
\begin{equation}\mathcal{G}_{IJ}\rightarrow
  \widetilde{\mathcal{G}}_{IJ}:=\Omega^2\mathcal{G}_{IJ}.  
\end{equation}
We can thus interpret minisuperspace as a conformal manifold
$(\mathcal{M},[ \mathcal{G}_{IJ}\dd q^I \otimes \dd q^J])$, that is,
as a manifold equipped with an equivalence class of metrics, 
\bdm
[ \mathcal{G}_{IJ}\dd q^I \otimes \dd q^J]=\left\{ 
\Omega^2 \mathcal{G}_{IJ}\dd q^I \otimes \dd q^J \mid 
\Omega:\mathcal{M}\rightarrow \mathbb{R}_+
\right\}.
\edm
Objects of interest on such a manifold are conformally
covariant objects, for example tensors that transform like  
\begin{equation}
\mathcal{T}\rightarrow\widetilde{\mathcal{T}}=\Omega^{k}\mathcal{T}
\end{equation} under Weyl transformations. We call $k$ the conformal
weight of $\mathcal{T}$ and denote it by $\ww (\mathcal{T})=k$.
Because of the conformal nature of configuration space, a spacetime
which satisfies Einstein's equations can, in fact, be regarded as a
sheaf of geodesics on this space \cite{Witt70}. 

In geometrodynamics, quantization is performed formally by replacing the canonical
momenta according to the rule $ p_I \rightarrow -\ii\hbar
\dif{}{{q^I}}{}$
and substituting these expressions into the Hamiltonian constraint
\eqref{constraint} \cite{OUP}. 
This procedure leads to the minisuperspace Wheeler-DeWitt equation
\begin{equation}
\hat{\mathcal{H}}\Psi=0 \ , \quad \text{with}\quad
\hat{\mathcal{H}}=-\frac{\hbar^2}{2}\ `` \
\mathcal{G}^{IJ}\dif{}{q_I}{}\dif{}{q_J}{} \ " + \mathcal{V}, 
\end{equation}
where the quotation marks indicate the need for choosing an appropriate
factor ordering.

The underlying conformal structure of minisuperspace motivates us to
choose a factor ordering that makes the Wheeler-DeWitt equation
conformally covariant. Following the discussion by Misner (\cite{magic},
p.~462),\footnote{See also \cite{Halliwell}.} this is achieved by
\begin{equation}
\left[ -\frac{\hbar^2}{2}\left(\square - \xi \mathcal{R} \right) + \mathcal{V}
\right]\Psi=0,
\label{eq:conformal_WDW}
\end{equation}
where $\mathcal{R}$ denotes the Ricci scalar constructed from
$\mathcal{G}_{IJ}$ and $\xi=\frac{d-2}{4(d-1)}$, with $d=\dim
(\mathcal{M})$.  
If we, in addition, impose the weights 
$\ww (\Psi)=-(d-2)/2$ and $\ww (\mathcal{V})=-2$, the Wheeler-DeWitt equation
(\ref{eq:conformal_WDW}) is indeed conformally covariant. The operator
$\square - \xi \mathcal{R}$ is called the conformal Laplacian or
Yamabe operator. It was shown that, given a compact Riemannian manifold of
dimension $d\geq 3$, one can find a metric conformal to ${\mathcal
  G}_{IJ}$ with constant scalar curvature \cite{LP87}.

Let us now turn to the discussion of the criteria for singularity avoidance.
As mentioned in the Introduction,
the first one goes back to DeWitt 
\cite{DeWitt}, who suggested to take $\Psi\rightarrow 0$
near the region of the classical singularity as a
sufficient criterion for quantum avoidance. In a heuristic
  sense, this corresponds to ``probability zero'' for the singularity.
Application of this criterion is based on the
idea that the (square of) the wave function is related to probability, as
is the case in quantum mechanics. In quantum gravity, this is far from
clear \cite{OUP}. The main reason is the absence of an external time
parameter in the Wheeler-DeWitt equation. Only in the semiclassical
(Born-Oppenheimer) approximation, where an approximate time parameter
emerges, can one impose the usual probability
interpretation in a straightforward manner. Nevertheless, we shall
stick heuristically to this 
idea also in the full theory. Peaks of the wave function have often
been interpreted as giving predictions in cosmology; see, for example,
\cite{BKKS10} and the references therein.

In the semiclassical limit with only one WKB component, an
interpretation using probabilties in
minisuperspace was suggested in \cite{HP86}; see also
\cite{Halliwell90}, pp.~186--190. Because the DeWitt criterion
rests on the heuristic notion of a probability, 
we find it appropriate to
include in this section  some remarks on the formulation of
this proposal in the language of conformal minisuperspace.

Let us consider solutions of the Wheeler-DeWitt equation in the WKB
approximation given by 
$\Psi \approx\sqrt{D}\ \ee^{{\ii}/\hbar} S$,
where $S$ is a solution to the Hamilton-Jacobi equation
\begin{equation}
\frac{1}{2}\mathcal{G}^{IJ}\left(\partial_I S \right)\left(\partial_J
  S \right) + \mathcal{V}=0 ,
\end{equation}
 and $D$ is the van Vleck factor which satisfies the linear transport equation
\begin{equation}
\mathcal{G}^{IJ}\left(\partial_I S \right)\partial_J D = -(\square S)D .
\end{equation}
Let now  $A \subseteq \mathcal{M}$ be a region in minisuperspace and
$B$ a thin `pencil' drawn out by the classical solutions, that is,
integral curves of the vector field  $\mathcal{G}^{IJ}(\partial_I
S)\partial_J$. 
It was shown in \cite{HP86} that
\begin{equation}
\int_{A \cap B} \star |\Psi|^2 
\approx \int_{A \cap B} \star D  
\approx F(B)
\int N \dd  t  , 
\label{eq:Hawking_Page_formula}
\end{equation}
where
\bdm
F(B):=\int_{\Sigma\cap B} D(\partial_I S)  \star \dd q^I,
\edm
 is the conformally invariant and conserved flux through a
hypersurface $\Sigma$ crossing the pencil $B$.\footnote{It is assumed
  that the hypersurface is chosen such as to cross each classical
  trajectory in the pencil only once \cite{Vilenkin89,HP86,Halliwell90}.}  
The contribution of $B$ to $\int_{A \cap B} \star |\Psi|^2 $ is
therefore proportional to the coordinate-time that the classical
solutions filling out the pencil $B$ spend in the region $A$. 
Note that $\text{w}(\star D)=\text{w}(\star |\Psi|^2)=2$. This
reflects the fact that the integral $\int N \dd  t $ on the right-hand side
of \eqref{eq:Hawking_Page_formula} depends on the representation of
the lapse which we choose before the quantization of the Hamiltonian
constraint in order to obtain the Wheeler-DeWitt equation
\eqref{eq:conformal_WDW}. 
In this sense, a conformal rescaling $\mathcal{G}_{IJ}\rightarrow
\Omega^2 \mathcal{G}_{IJ}$, $\Psi \rightarrow \Omega^{\ww (\Psi)}\Psi$  
 corresponds to a time reparametrization at the quantum level.
Equation \eqref{eq:Hawking_Page_formula} 
can also help us to interpret the behavior of wave 
packets in regions of minisuperspace where the WKB approximation is
valid. 

The DeWitt criterion was successfully applied to cosmological models in a
series of recent 
papers; see, for example, \cite{Bouhmadi16} and references therein. 
These examples deal mostly with two-dimensional mini\-superspaces where
$\ww (\Psi)=0$ and the usual Laplace-Beltrami operator coincides with
the conformal 
Laplacian.
In dimensions $d\geq 3$, however, the DeWitt criterion is
not conformally invariant. Moreover, there does not seem to be a
  privileged representative of the wave function for the imposition of
  the criterion. For the reasons mentioned above, we seek here a
generalization of the DeWitt criterion to 
guarantee its conformal invariance.

 This  leads us to consider
conformally invariant objects constructed from $\Psi$. We first note
that we can define 
a density of conformal weight 0 by 
\begin{equation}
\star |\Psi|^\frac{2d}{d-2}=|\Psi|^\frac{2d}{d-2} \text{dvol},
\end{equation}
where $\text{dvol}$ contains the square root of the (absolute value
of the) determinant of the DeWitt metric,
and $\star$ denotes the Hodge star.
Moreover, we address the Klein-Gordon current defined by
\begin{equation} 
\mathbf{J}[\Psi_1,\Psi_2]=\frac{1}{2\ii}\star\left(\Psi_1^* \dd \Psi_2
  -\Psi_2 \dd \Psi_1^* \right),  
\end{equation}
which is a ($d-1$)$-$form with conformal weight 0.
These definitions allow us to propose the following two criteria.

\theoremstyle{criterion1}
\begin{criterion1}
A singularity is said to be avoided if
$\star|\Psi|^\frac{2d}{d-2}\rightarrow 0$ in the vicinity of the
singularity. 
\end{criterion1}

This is the conformally invariant version of the DeWitt criterion
\cite{DeWitt}. 

\theoremstyle{criterion2}
\begin{criterion2}
A singularity is said to be avoided if
$\mathbf{J}[\Psi,\Psi]\rightarrow 0$ in the vicinity of the
singularity.  
\end{criterion2}

Another criterion, which was introduced in the discussion of the
quantum fate of the big-rip singularity in \cite{DKS}, is the
following: 

\theoremstyle{criterion3}
\begin{criterion3}
A wave packet is said to avoid the singularity if it spreads in the
vicinity of the singularity. 
\end{criterion3}

The spreading of wave packets indicates the breakdown of the semiclassical
approximation. Classical cosmology and in particular the classical
singularity theorems then cease to hold. The notion of a
  classical spacetime can no longer be applied, which leads to the end of
  classical predictability before reaching the singularity.
This criterion is fulfilled, for example,
in the big-rip case studied in \cite{DKS}. 

We note that the second criterion suffers from the problem that it is
not applicable in the case of real wave functions, which often arise
as solutions to the (real) Wheeler-DeWitt equation. An
  example of a real wave function is the no-boundary
  (Hartle-Hawking) state. In contrast to this, the ``tunneling wave
  function'' is complex, and criterion~2 can be applied.\footnote{See,
    for example, \cite{OUP,Calcagni} for a detailed discussion of
    boundary conditions.}

The Klein-Gordon
flux is not positive definite and can thus in general not be
interpreted as a probability flux. Exceptions are situations where
only one WKB branch is present; this has led to the proposal that the
Klein-Gordon current  only be applied to such cases
\cite{Vilenkin89}. The case of one WKB wave function can also be
  interpreted as a decohered branch of a real wave function.
In the following, we shall thus mainly concentrate
on the first criterion, which is the natural
generalization of the DeWitt criterion to higher-dimensional
minisuperspaces.

\section{Kasner solution}

The vacuum Bianchi~I (Kasner) solution can be written in the form
\begin{align}
&\dd s^2 = -\dd t^2 + t^{2p_x}\dd x^2+ t^{2p_y}\dd y^2+ t^{2p_z}\dd z^2 
\quad \text{with}\\
&p_x^2+p_y^2+p_z^2=1 \quad \text{and} \quad p_x+p_y+p_z=1 .
\end{align}
The constraints on $p_x$, $p_y$ and $p_z$ define the so-called Kasner
sphere and Kasner plane, respectively. 
The physical solutions lie on their intersection, which represents a circle
in the $(p_x,p_y,p_z)$ space. 
The nature of the singularity depends on the value of the coefficients
$p_x,p_y,p_z $. 
If one of them is equal to 1, the Kasner solution will become the Milne
universe, which is diffeomorphic to slices of Minkowski
spacetime. The singularity is then only a coordinate  singularity. For
all other values, the singularity is physical, which is indicated by the
divergence of the Kretschmann invariant, 
\begin{equation}
R_{\mu\nu\lambda \sigma}R^{\mu\nu\lambda \sigma} = C_{\mu\nu\lambda
  \sigma}C^{\mu\nu\lambda \sigma}  +2 R_{\mu\nu}R^{\mu\nu}-\frac13R^2.
\end{equation}
Here, the singularity is a big bang (or big crunch) singularity.
Since $R_{\mu \nu}=0$, the curvature singularity is a pure Weyl singularity. 
If we use the common parametrization of the Kasner circle,
\begin{equation}
\begin{aligned}
p_x=-\frac{u}{1+u+u^2}, \
p_y=\frac{1+u}{1+u+u^2}, \
p_z = \frac{u(1+u)}{1+u+u^2},
\end{aligned}
\end{equation}
with $u\in (-\infty,\infty)$, we find that
\begin{equation}
C_{\mu\nu\lambda
  \sigma}C^{\mu\nu\lambda \sigma}=\frac{16(1+u)^2 u^2}{(1+u+u^2)t^4} \ .
\end{equation}
In order to obtain a Hamiltonian for the model, we employ the general symmetry
reduced ansatz 
\bdm
\begin{aligned}
&\dd s^2 = -N^2 \dd t^2 +a_x^2 \dd x^2  +a_y^2 \dd y^2 + a_z^2 \dd z^2,
 \quad \text{where} \\  
& a_x= a\ee^{ \beta_+ + \sqrt{3}\beta_-},
a_y=a\ee^{ \beta_+ - \sqrt{3}\beta_-}  
\ \text{and} \ a_z=a\ee^{ - 2 \beta_+},
\end{aligned}
\edm
with the scale factor $a=(a_x a_y a_z)^{1/3}=: \ee^{\alpha}$,
whose third power describes the volume (which expands as $a^3\propto t$), and
with the anisotropy factors $\beta_{\pm}$, 
which describe the shape of the universe.
Note that the scale factor is chosen here to be dimensionless; the
physical length dimension is in the coordinates $x$, $y$, and $z$. 

The symmetry reduced Einstein-Hilbert action takes the form 
\begin{equation} 
S_{\rm EH}
=
\frac{1}{2}\int \dd t \ \frac{\ee^{3\alpha}}{N} \left( 
-\dot{\alpha}^2
+
\dot{\beta}_+^2
+
\dot{\beta}_-^2 \right) .
\label{eq:S_EH}
\end{equation}
The  Hamiltonian  obtained after the usual Legendre transform reads
\begin{equation} 
H=N\mathcal{H}=\frac{N \ee^{-3\alpha}}{2}\left(
-p_{\alpha}^2 + p_{+}^2 + p_{-}^2  
\right) .
\label{eq:H_BI}
\end{equation}
Choosing for the lapse function the value $N=\ee^{3\alpha}$, it becomes
clear that the 
Hamiltonian is equivalent to the Hamiltonian of a free relativistic
particle in $2+1$ dimensions. 
We conclude that the solutions represent straight lines in 
minisuperspace, which can be parametrized as follows:
\begin{equation}
\beta_\pm = \frac{p_\pm}{\sqrt{p_+^2 + p_-^2}} \alpha + C_{\pm},
\label{EQ_BETA_KASNER}
\end{equation}
with $C_{\pm}\in \mathbb{R}$ arbitrary constants. The approach to the
singularity is called velocity term dominated (VTD); see, for
example, \cite{Berger}. This terminology refers to the dominance of
the kinetic over the potential terms, which is trivially
fulfilled here (absence of potential). 

The DeWitt metric on $\mathcal{M}$ is given by
\begin{equation} 
\mathcal{G}_{IJ}\ \dd q^I \otimes \dd q^J=\ee^{3\alpha}\left(-\dd \alpha^2 + \dd \beta_+^2 +\dd \beta_-^2 \right),
\end{equation}
from which we obtain for the Ricci scalar on
$\mathcal{M}$ the value $\mathcal{R}=\frac{9}{2}\ee^{-3\alpha}$. The
Wheeler-DeWitt equation now reads
\bdm
\frac{\hbar^2\ee^{-3\alpha}}{2}\left[
\dif{}{\alpha}{2}+2f\dif{}{\alpha}{}
 -
\dif{}{\beta_+}{2}
 -
\dif{}{\beta_-}{2}+{\xi}\mathcal{R}\ee^{3\alpha}
\right]\Psi=0,
\edm
where the numbers $f$ and $\xi$ parametrize a family of operator
orderings. 
After the transformation
$\Psi\rightarrow\widetilde{\Psi}=\ee^{f\alpha}\Psi$ we obtain 
\begin{equation} 
\left[
-\dif{}{\alpha}{2}
+\dif{}{\beta_+}{2}
+\dif{}{\beta_-}{2}
+ f^2-\frac{9}{2}{\xi} \right]\widetilde{\Psi}=0.
\end{equation}
The conformal factor ordering is obtained by setting $f=3/4$ and
$\xi=1/8$. We then get 
\begin{equation}
  \label{flat}
\left[
-\dif{}{\alpha}{2}
+\dif{}{\beta_+}{2}
+\dif{}{\beta_-}{2}
\right]\widetilde{\Psi}=0.
\end{equation}
In the conformal factor ordering the Wheeler-DeWitt equation is thus
identical to the classical wave equation in $d=1+2$ dimensions. 
Note that the DeWitt metric is flat in this representation, such that
  criterion~1  above is equivalent to the DeWitt criterion $\widetilde{\Psi}\rightarrow
  0$ as applied in earlier papers; see, for example,
  \cite{Bouhmadi16}.

Let us now turn to the formulation of the criteria for singularity avoidance.
There, the minisuperspace dimension $d$ will be crucial.
Solutions to the free wave equation in $1+1$ dimension can
propagate only into two directions. Wave packets are not subject to
spreading and their amplitudes do not decay.   
In higher dimensions, however, the wave can propagate into infinitely
many directions. This leads to a spreading and a resulting decay of
the amplitude of the wave.  
The above statement can be made more precise in the form of 
{\em decay rate estimates}.  

\begin{figure}[!ht]
	\centering
	\includegraphics[width=9.0cm,angle=0]{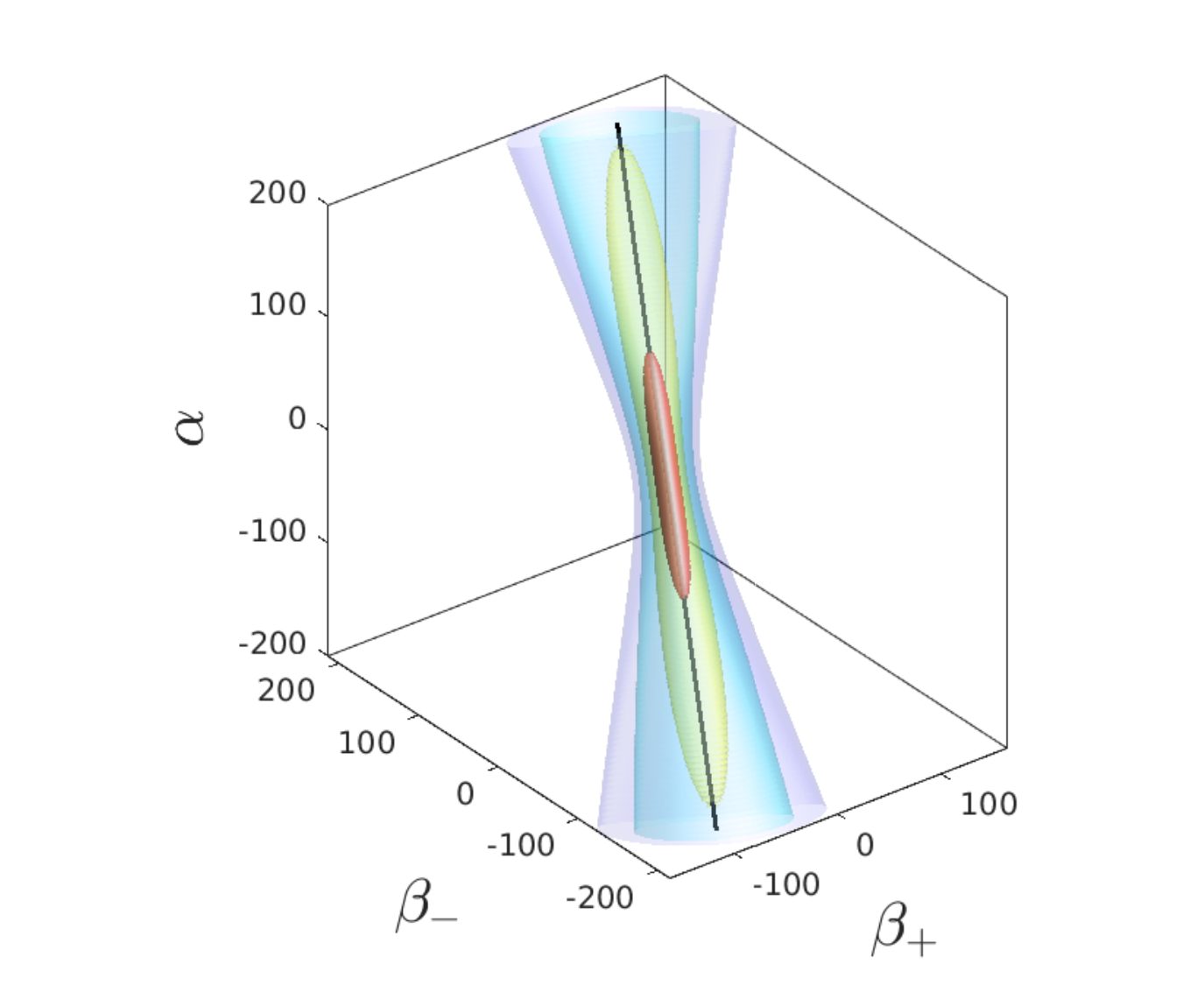} 
	\caption{Plot of equipotential surfaces of $|\Psi|$ for a wave
          packet $\Psi$ solving the Wheeler-DeWitt equation
          \eqref{flat}. The thin black line is the classical
          trajectory.}  
	\label{fig:Kasner_wavepacket} 
\end{figure}

In $d>2$ dimensions, we can apply the following decay rate estimate
(see e.g. \cite{Klainerman85}):
Let $\Phi$ be a solution to the initial value problem
\bdm
\begin{cases}
\left[\dif{}{t}{2}-\sum\limits_{i=1}^{d-1}\dif{}{x_i}{2} \right]\Phi 
=0 
 \quad (t,x_1 , ... ,x_{d-1})\in \mathbb{R}_+ \times \mathbb{R}^{d-1} 
 \\   \quad
(\Phi,\partial_t \Phi)\mid_{t=0} 
= 
(f,g) 
\end{cases} ,
\edm
where $f$ and $g$ are smooth functions
$\mathbb{R}^{d-1}\rightarrow\mathbb{R}$ with compact support. Then
there exist   $C_{1/2}>0$ such that  
\bdm
\begin{aligned}
|\Phi(t,\mathbf{x})|\leq C_1|t|^{-\frac{d-2}{2}}  \  &\text{and}
\  
|\partial_i \Phi(t,\mathbf{x})|\leq C_2| t |^{-\frac{d-2}{2}} .
\end{aligned}
\edm
For the situation in question it follows that such wave packets
satisfy the above criteria 1, 2, and 3 for singularity avoidance, that
is we have, 
\begin{equation}
  \label{avoidance-kasner}
\mathbf{J}[\Psi, \Psi] \rightarrow 0 \quad \text{and}\quad
\star|\Psi|^{6}\rightarrow 0 \quad \text{as} \ \  \alpha\rightarrow
\pm \infty .
\end{equation} 
This is caused by the spreading of the wave packet when approaching the
region of the classical singularity.
The singularity is thus avoided by all criteria.
  Since the Wheeler-DeWitt equation \eqref{flat} is symmetric with
  respect to $\alpha\to-\infty$, the same conclusion holds for
  $\alpha\to+\infty$. What does this mean? In quantum mechanics, such
  a spreading can also occur, for example in the case of a free
  particle. But there one has unitarity with respect to external time
  $t$ and the standard scalar product. In quantum cosmology,
 there is no consensus about the choice of inner product. If one used
 a norm motivated by the conformally invariant DeWitt
 criterion, that is, an integral over $\beta_+$ and $\beta_-$ of
 $\star|\Psi|^{6}$, this is not conserved in $\alpha$; one can
 even estimate that the integral goes to zero for $\alpha\to+\infty$,
 so the situation is very different from the quantum mechanical free
 particle: there is {\em no} unitarity with respect to the timelike
 variable $\alpha$. In the spirit of the DeWitt criterion, one could
 call this a quantum avoidance of the late-time evolution.
 Figure~\ref{fig:Kasner_wavepacket} displays the behavior of the wave
 packet for $\alpha\to-\infty$ and $\alpha\to+\infty$.
 
This is an example where quantum effects are not restricted to small
 scales of Planck size. Because the superposition principle is
 universally valid in quantum cosmology, quantum effects can arise in
 principle at any scale. One example is
 the turning point of a classically recollapsing universe \cite{KZ95},
 where destructive interference has to occur in order to guarantee a
 recollapsing wave packet.

One could argue that a natural inner product for the Wheeler-DeWitt
 equation is the Klein-Gordon inner product, which provides unitarity
 with respect to $\alpha$. The vanishing of the
 Klein-Gordon current corresponds to our criterion~2 of singularity
 avoidance and is 
 fulfilled in the present case, both for $\alpha\to-\infty$ and
 $\alpha\to+\infty$, see (\ref{avoidance-kasner}).

 A somewhat different approach to the quantization of Bianchi~I vacuum
 models with and without cosmological constant was developed in
 \cite{CGP02}. There, the dynamics was reduced such that the wave
 function depends only on one degree of freedom, the determinant of
 the scale factor $a$. How a singularity avoidance in this approach
 relates to the singularity avoidance discussed here, is an
 interesting question that is beyond the scope of our investigation.

In the next section, we will investigate if and how the situation
changes if matter is added to the model.

\section{Bianchi I model with an effective matter potential}

In this section, we treat matter in a phenomenological way. The
representation of matter by a dynamical scalar field $\phi$ is
relegated to the next section. In anisotropic models, anisotropic
pressures can be used, but we address for simplicity the case of a
barotropic fluid. 

A hypersurface orthogonal (non-tilted) barotropic fluid with an
equation of state $p=w\rho$ and 
$\rho\propto a^{-3(1+w)}$ can be modelled by adding an effective
matter potential of the form $\mathcal{V}(\alpha)=N \mathcal{V}_0
\ee^{-3(1+w) \alpha} \propto \rho$ to the Einstein-Hilbert action
(\ref{eq:S_EH}), with $\mathcal{V}_0 > 0$ being
constant. The full action then reads 
\bdm
S
=
\int \dd t \ \ee^{3\alpha} \left( \frac{-\dot{\alpha}^2
+
\dot{\beta}_+^2
+
\dot{\beta}_-^2}{2N}
 - N\mathcal{V}_0\ee^{-3(1+w)\alpha} \right).
\edm
One recognizes that the introduction of matter introduces an
  asymmetry with respect to $\alpha$.
We restrict our discussion to $w<1$, which excludes the case of a
stiff matter fluid. The important cases of a cosmological constant
($w=-1$), dust ($w=0$), and radiation ($w=1/3$) are included.  
The null and weak energy conditions are satisfied for $w\geq -1$,
  while the strong energy condition and the dominant energy condition
  require $w\geq -\frac{1}{3}$ and $-1\leq w \leq 1$, respectively. 

\begin{figure}[!ht] \centering
  \begin{subfigure}[b]{0.48\textwidth}
    \centering
    \includegraphics[width=\textwidth]{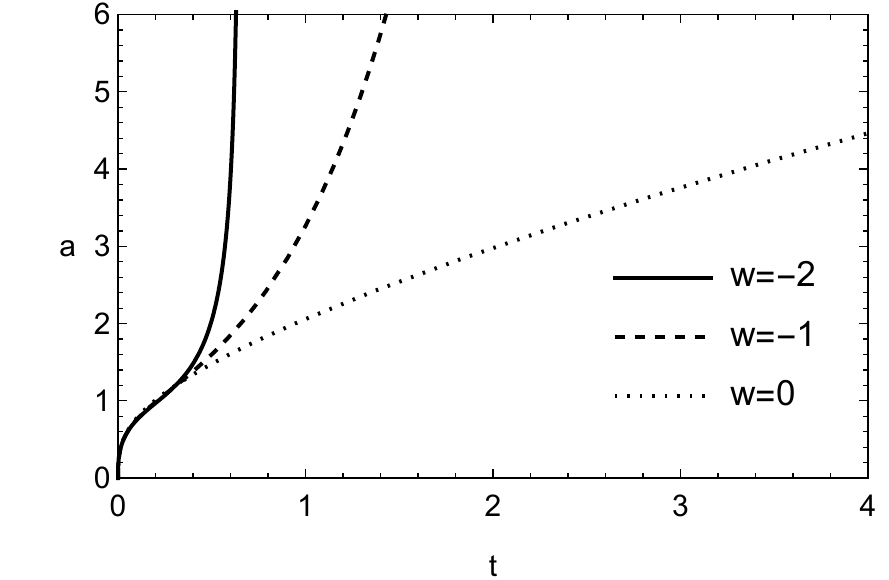}
    \caption{Scale factor $a(t)$} 
    \label{fig:effpot_scale}
  \end{subfigure}%
  \hspace{0.5cm}
  \begin{subfigure}[b]{0.48\textwidth}
    \centering
    \includegraphics[width=\textwidth]{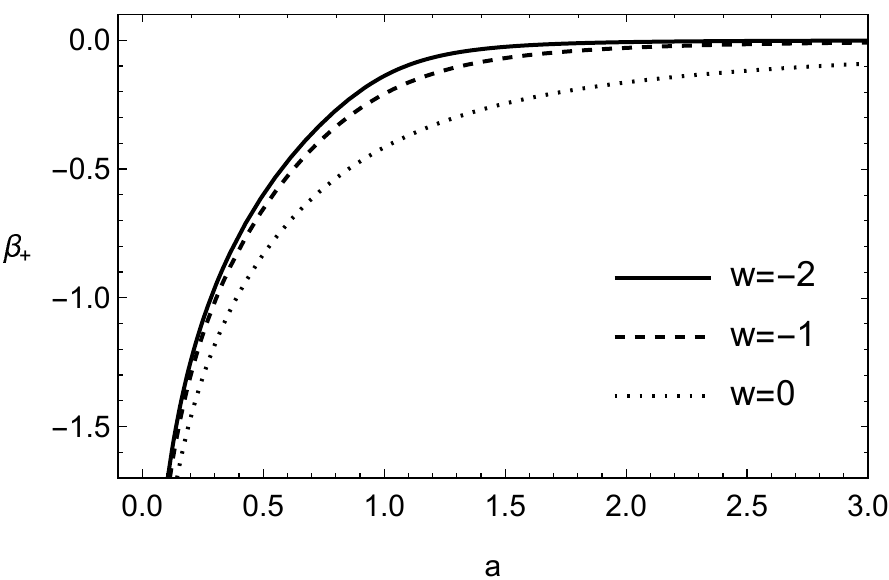}
    \caption{Anisotropy factor $\beta_+(a)$} 
    \label{fig:effpot_shape}
  \end{subfigure}
  \caption{Scale factor $a(t)$ and anisotropy factor $\beta_+(a)$ for
    different $w$ and $p_+ = p_- = \mathcal{V}_0 = 1$. One
      recognizes isotropization for large universes.}
  \label{fig:effpot_classical}
\end{figure}

The variables $\beta_\pm$ are cyclic and we call their
conserved conjugate momenta
$p_\pm$; cf. \eqref{eq:H_BI}.  Variation of the Lagrangian with respect to $N$ 
leads to 
\begin{equation}
\dot{a}^2 = N^2 \left( p_+^2 + p_-^2 + 2\mathcal{V}_0 a^k \right) a^{-4},
\label{DIFF_EQ_A}
\end{equation}
where $k:=3(1-w)$.
We assume that $p_+^2 + p_-^2\neq 0$ and choose the comoving gauge
$N=1$. Equation \eqref{DIFF_EQ_A} is then solved  by 
\begin{equation} 
\label{hypergeometric}
t = \frac{a^3}{3\sqrt{p_+^2 + p_-^2}} {}_2F_1 \left[ \frac{1}{2},
  \frac{3}{k}; 1+\frac{3}{k}; -\frac{2\mathcal{V}_0}{p_+^2 + p_-^2}
  a^k \right], 
\end{equation}
where ${}_2F_1[a,b;c;z]$ is the hypergeometric function. The scale factor
$a(t)$ is
shown for different $w$ in Fig.~\ref{fig:effpot_scale}. 

For small $a$,
the hypergeometric function asymptotically equals 1, and we get for
$a\to 0$: 
\begin{equation}
t \sim \frac{a^3}{3\sqrt{p_+^2 + p_-^2}}. 
\end{equation}
Thus the universe starts with a big bang at $t=0$, independent of the
value for the barotropic index $w$. For large $a$ and $w\neq -1$, 
the hypergeometric function can be simplified, too, and one gets from
\eqref{hypergeometric} in the limit $a\to\infty$:
\begin{equation} 
t \sim \sqrt{\frac{2}{\mathcal{V}_0}} \frac{1}{6-k} a^{(6-k)/2} + t_{*} .
\label{T_OF_A_LARGE_LIMIT}
\end{equation} 
For $k<6$ ($w>-1$), the universe expands infinitely, whereas in the
phantom case, that is for $k>6$ ($w<-1$), the universe becomes infinitely
large already at $t=t_{*}$ and ends with a big rip. We note that
(\ref{T_OF_A_LARGE_LIMIT}) is the full solution for the flat FLRW
case: for $k<6$ (non-phantom case), there is a big bang, but
for $k>6$ (phantom case) there is no past 
singularity. Therefore one can say that the anisotropy introduces  the
past singularity, leading to a model with big bang {\em and} big rip. 

For the anisotropy factors one has
\begin{equation}
\beta_\pm = \frac{1}{k} \frac{p_\pm}{\sqrt{p_+^2 + p_-^2}} \log\left|
  \frac{1 - \sqrt{1+ \frac{2\mathcal{V}_0}{p_+^2 + p_-^2} a^k}}{1 +
    \sqrt{1+ \frac{2\mathcal{V}_0}{p_+^2 + p_-^2} a^k}} \right|; 
\label{EQ_BETA_OF_A}
\end{equation}
they become constant for large $a$, see also
Fig.~\ref{fig:effpot_shape}. Thus in contrast to the vacuum solution,
this universe isotropizes for late times. For small $a$, the asymptotic
behavior corresponds to (\ref{EQ_BETA_KASNER}), which is again independent of
the matter content. 
This property is sometimes called ``matter doesn't matter''.
Since the Kasner behavior is recovered in the limit $a\rightarrow 0$, this
approach to the singularity is referred to as asymptotically velocity
term dominated (AVTD) \cite{Berger}. 

We now turn to the quantum version of these models. 
The Wheeler-DeWitt equation reads
\begin{equation}
\left[
\dif{}{\alpha}{2}
 -
\dif{}{\beta_+}{2}
 -
\dif{}{\beta_-}{2}
+\mathcal{V}_0 \ee^{k\alpha}
\right]\Psi=0,
\end{equation}
where  we have set $\hbar=1$ and skipped the tilde over the wave function.
The solutions can be written in the form
\bdm
\begin{aligned}
\Psi \left(
\alpha,
\beta_+,
\beta_- 
\right) &=
\\
\sum\limits_{\sigma=\pm}
\int_{\mathbb{R}^2} & \dd p_+ 
\dd p_- \ 
\mathcal{A}_\sigma\left(p_+,p_- \right) 
\psi_{p_+,p_-}^\sigma \left(
\alpha,
\beta_+,
\beta_- 
\right),
\label{eq:eff_pot_wacepacket}
\end{aligned}
\edm 
with the mode functions given by
\begin{equation}
\begin{aligned}
& \psi^{\pm}_{p_+,p_-}
\left(\alpha,\beta_+,\beta_-\right)
=  \ \ee^{-\ii p_+ \beta_+ - \ii p_- \beta_-}
\\
&  \times
c_{p_+,p_-}^{\pm}
J_{\pm \frac{2\ii}{k}\sqrt{p_+^2+p_-^2}}\left(
\frac{2}{k} \sqrt{\mathcal{V}_0} \ee^{k\alpha/2}
\right),\\
& c_{p_+,p_-}^{\pm}
:=   \
\Gamma\left(1\pm \frac{2\ii}{k}\sqrt{p_+^2+p_-^2} \right) \left(
  \frac{\sqrt{\mathcal{V}_0}}{k} \right)^{\mp 2\ii \sqrt{p_+^2 +
    p_-^2}/k} , 
\end{aligned} 
\label{eq:mode_functions}
\end{equation}
where $J_\nu(z)$ and $\Gamma(z)$ denote the Bessel function of the
first kind and the gamma function, respectively. 
Let us now investigate the asymptotic forms of the wave packet. 
In the limit  $\alpha\rightarrow -\infty$ we can approximate the mode functions by 
\bdm
\psi^{\pm}_{p_+,p_-}
\left(\alpha,\beta_+,\beta_-\right) 
= \ee ^{\pm\ii \sqrt{p_+^2+p_-^2} \ \alpha -\ii p_+\beta_+ - \ii p_-\beta_-} +
\mathcal{O}\left( 
 \ee^{k\alpha}
\right),
\edm
which is independent of $k$.
We conclude that the quantum Kasner behavior is recovered in this
limit (which follows as a solution of \eqref{flat}). 

The discussion of the limit  $\alpha\rightarrow \infty$  is slightly
more complicated, but it turns out that a discussion of the mode
functions in the WKB approximation  
\begin{equation}
\psi
\approx  {\sqrt{D}}
\exp\left(\ii S\right)
\end{equation}
will be sufficient.
A solution to the Hamilton-Jacobi equation is given by
\ben
& & S_{p_+,p_-}(\alpha, \beta_+,\beta_-)= \pm
\left(  
\frac{2}{k} \sqrt{p_+^2+p_-^2+ \mathcal{V}_0 \ee^{k\alpha}}\right.\\
& & \ \ \ \left.+ \frac{1}{k} \sqrt{p_+^2 + p_-^2} \log\left|
  \frac{1-\sqrt{1+\frac{\mathcal{V}_0}{p_+^2 + p_-^2}
      \ee^{k\alpha}}}{1+\sqrt{1+\frac{\mathcal{V}_0}{p_+^2 + p_-^2}
      \ee^{k\alpha}}} \right| \right)\\
& & \ \ \ -p_+\beta_+ - p_-\beta_- . 
\een
The corresponding van Vleck factor reads
\begin{equation}
D_{p_+,p_-}(\alpha)=
\frac{1}{\sqrt{ p_+^2 + p_-^2 + \mathcal{V}_0 \ee^{k\alpha}}}.
\label{eq:van_Vleck_factor}
\end{equation}
If we introduce the functions
\bdm
\begin{aligned}
\mathcal{B}_{+}(p_+,p_-)  = &
\sqrt{\frac{k}{8\pi}}(1-\ii)
\left[
c^+_{p_+,p_-}
\ee^{\frac{\pi}{k}\sqrt{p_+^2+p_-^2}} 
\mathcal{A}_{+}(p_+,p_-)
\right. 
\\
&\left.
+
c^-_{p_+,p_-}
\ee^{-\frac{\pi}{k}\sqrt{p_+^2+p_-^2}} 
\mathcal{A}_{-}(p_+,p_-)
\right] ,
\\
\mathcal{B}_{-}(p_+,p_-)  = &
\sqrt{\frac{k}{8\pi}}(1+\ii)
\left[
c^+_{p_+,p_-}
\ee^{-\frac{\pi}{k}\sqrt{p_+^2+p_-^2}} 
\mathcal{A}_{+}(p_+,p_-)
\right. 
\\
&\left.
+
c^-_{p_+,p_-}
\ee^{\frac{\pi}{k}\sqrt{p_+^2+p_-^2}} 
\mathcal{A}_{-}(p_+,p_-)
\right],
\end{aligned}
\edm
then the approximate wave packet with these coefficients,
\begin{equation}
\sum\limits_{\sigma=\pm}\int_{\mathbb{R}^2}\dd p_+ \dd p_-\
\mathcal{B}_{\sigma} { \sqrt{D}} 
\exp\left(\sigma \ii S\right),
\end{equation}
matches the exact  wave packet for large $\alpha$  at the leading order. 
This follows from the asymptotic expansion of the exact mode functions
and an approximation of the WKB modes of the form 
\bdm
\psi \approx \frac{1}{\sqrt[4]{\mathcal{V}_0}}
\ee^{-k\alpha/4} \exp\left[\pm\ii \left( \frac{2}{k}
    \sqrt{\mathcal{V}_0} \ee^{k\alpha/2} \right) \right]. 
\edm
Then one has
\begin{equation}
\begin{aligned}
& \Psi(\alpha,\beta_+,\beta_-)   
  \approx 
 \frac{\ee^{-\frac{k}{4}\alpha}}{\sqrt[4]{\mathcal{V}_0}}  \sum\limits_{\sigma=\pm}
\exp
\left(\sigma
\frac{2\ii}{k} \sqrt{\mathcal{V}_0}
\ee^{\frac{k}{2}\alpha}
\right)
\\
 \times &
\int_{\mathbb{R}^2} \dd p_+ 
\dd p_- \
{\mathcal{B}}_{\sigma}\left(p_+,p_- \right)\ee^{-\ii p_+ \beta_+ -\ii p_- \beta_-}.
\end{aligned} 
\label{eq:Psi_large_a}
\end{equation}
We can now draw a clear picture of the behavior of wave packets. In
the limit $\alpha\rightarrow-\infty$, we recover the quantum Kasner
behavior. 
Consequently, we expect a spreading with a resulting decay of
amplitudes. The behavior in the limit $\alpha\rightarrow\infty$ can
be inferred from (\ref{eq:Psi_large_a}):  
the term in the second line of this equation is just the
Fourier transform of $B_\sigma$ and is independent of $\alpha$. If,
for example, we choose $B_\sigma$ to be Gaussian, its Fourier transform
will be a Gaussian which is peaked around some particular values of $\beta_+$
and $\beta_-$. This strongly reflects the classical behavior of
isotropization.    
Most importantly, wave packets do not spread in the region where $\alpha$ is large.
The wave packet is modulated by a strongly oscillating factor and an
exponentially decaying factor. The exponentially decaying factor comes
from the van Vleck factor \eqref{eq:van_Vleck_factor} and can be interpreted
as arising from the particular representation of the wave
function.

The decay of the mode functions in this representation can be
intuitively understood by inspecting the Hawking-Page formula
\eqref{eq:Hawking_Page_formula}: The representation of the wave
function $\Psi$ we are working with is related to the gauge
$N=\ee^{3\alpha}$ by the corresponding  representation  of the DeWitt
metric. In this gauge, classical solutions reach  
$\alpha=\infty$ in a finite time $t$. Hence they spend less and less
time $t$ in the region of minisuperspace where $\alpha$ is large. In
this sense the decay of the density $\sqrt{-\mathcal{G}}D$ is implied
by \eqref{eq:Hawking_Page_formula}. 

\begin{figure}[!ht]
	\hspace{-0.7cm}
	\includegraphics[width=8.0cm,angle=0]{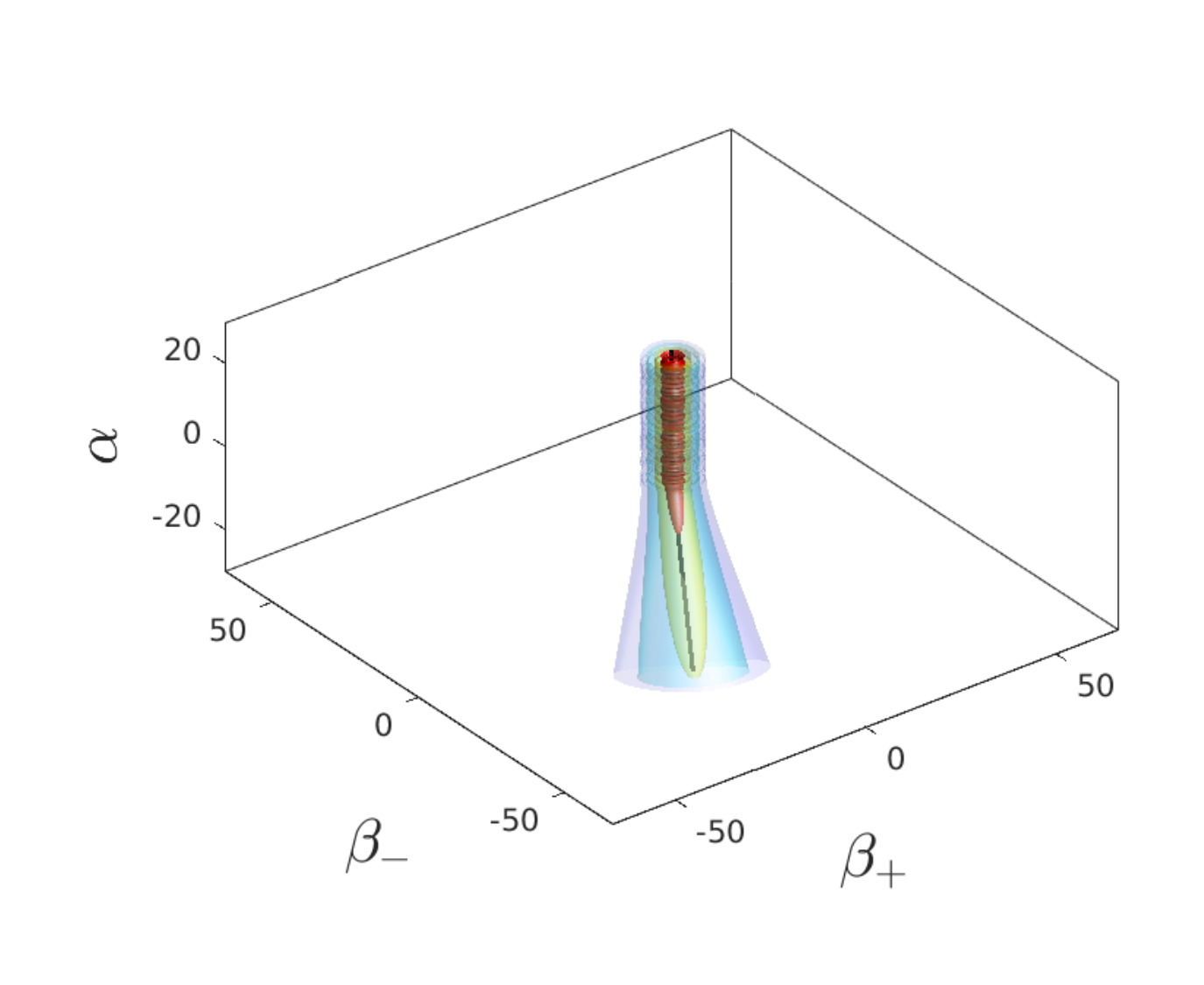} 
	\caption{Plot of the equipotential surfaces of the rescaled
          wavepacket $D_{\bar{p}_+,\bar{p}_-}^{-1/2}|\Psi|$ for the
          dust case ($k=3$). The amplitude   $\mathcal{A}_+$ was
          chosen to be a symmetric Gaussian peaked about some momenta
          $(p_+,p_-)=(\bar{p}_+,\bar{p}_-)$,  while $\mathcal{A}_- $
          was set to zero.  The thin black line is the corresponding
          classical trajectory. }   
	\label{fig:dust_wavepacket} 
\end{figure}

Figure~\ref{fig:dust_wavepacket} displays the behavior
  of the wave packet in the model with dust. The asymmetry compared
  to Fig.~\ref{fig:Kasner_wavepacket} is clearly visible.

For simplicity, we now set $\mathcal{B}_- = 0$. Then
the large-$\alpha$ limit of the Klein-Gordon current is given by 
\bdm
\begin{aligned}
\mathbf{J}[\Psi,\Psi]=  
& \left|\int_{\mathbb{R}^2} \dd p_+ 
\dd p_- \
{\mathcal{B}}_+
\ee^{-\ii p_+ \beta_+-\ii p_- \beta_-} \right|^2 \dd \beta_+ \wedge \dd \beta_- 
\\
& + \mathcal{O}\left(\ee^{-\frac{k}{4}\alpha}\right).
\end{aligned}
\edm
Up to leading order, the current only has an $\alpha$ component 
given by the Fourier transform of $B_+\left(p_+,p_- \right)$. If we
assume that $B_+$ is peaked at some particular values $p_+$ and $p_-
$, we will expect the Fourier transform of $B_+$ to be peaked at
some particular value of $\beta_+ $ and $\beta_-$. The current
thus reflects the classical behavior in the region where
$\alpha$ is large (in contrast to the vacuum Kasner case).  
We have, however,
$\quad \star |\Psi|^6 \rightarrow 0\quad$ as $\alpha\rightarrow\infty$. 
Note that the behavior is qualitatively independent of $w$,
that is, there is no difference between the cases $w \geq -1$ and $w <
-1$, although the  latter case leads to a big rip.  
The big rip is thus only avoided by criterion~1.

\section{ Bianchi~I model with a phantom field}

\begin{figure}[!tb] \centering
  \begin{subfigure}[b]{0.48\textwidth}
    \centering
    \includegraphics[width=\textwidth]{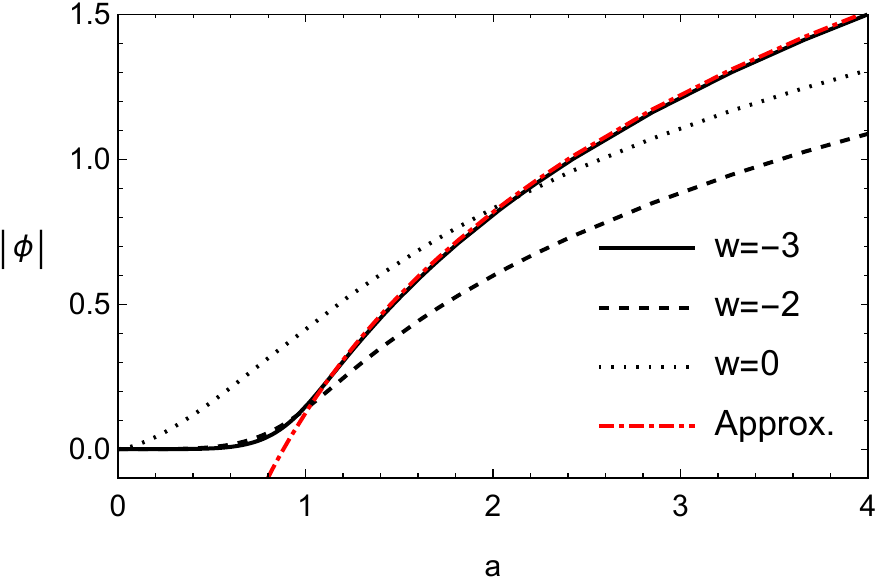}
    \caption{Scalar field $|\phi(a)|$} 
    \label{fig:scalar_phi}
  \end{subfigure}%
  \hspace{0.5cm}
  \begin{subfigure}[b]{0.48\textwidth}
    \centering
    \includegraphics[width=\textwidth]{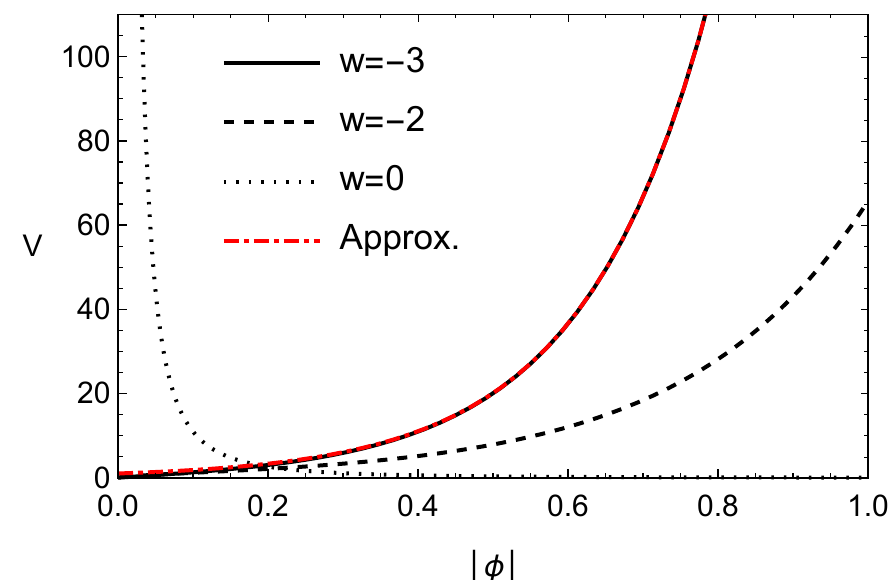}
    \caption{Scalar potential $V(\phi)$} 
    \label{fig:scalar_potential}
  \end{subfigure}
  \caption{Scalar field $|\phi|(a)$ and its potential $V(\phi)$ for
    different $w$ and $p_+ = p_- = \mathcal{V}_0 = 1$; the red curves
    correspond to the approximations (\ref{EQ_V_PHI_APPROX}).} 
  \label{fig:scalar_classical}
\end{figure}

In the previous section, we have added matter degrees of freedom
through an effective potential $\mathcal{V}(a)$. 
Here, we will instead implement the equation of state 
$p=w\rho$ by a scalar field $\phi$ with a potential $V(\phi)$, a
procedure described in \cite{Gorini04}. Matter degrees of
freedom are now dynamical.
The connection between the kinetic and potential terms and the
parameters $\rho$ and $p$ are as follows:
\begin{equation}
\dot{\phi}^2 = \frac{N^2}{l}(\rho+p) \quad, \quad V = \frac{1}{2}(\rho-p).
\label{EQ_PHI_V_RHO_P} 
\end{equation}
Note that $l=\pm 1$ depending on whether we consider normal or
phantom matter, respectively. Using these relations one finds the same
functions for 
$a(t)$ and $\beta_\pm(t)$ as in the previous section.
We use here $\kappa_\pm$ instead of
$p_\pm$, because these constants will be used in the construction of $V(\phi)$. 
Combining (\ref{EQ_PHI_V_RHO_P}), (\ref{DIFF_EQ_A}),
$p=w\rho$, and $\rho(a)$ we get for the classical solution in
configuration space, 
\begin{equation}
\phi(a) = \pm  \frac{1}{k} \sqrt{\frac{2}{3}l(6-k)} \arcsinh\left[
  \sqrt{\frac{2\rho_0}{\kappa_+^2 + \kappa_-^2}} a^{k/2} \right]  ,
\label{EQ_PHI_OF_A}
\end{equation}
see Fig.~\ref{fig:scalar_phi}. The scalar field vanishes like a
polynomial for small $a$ and diverges logarithmically for large $a$.  

Using the same equations as before we get for the potential
\begin{equation}
V = \rho_0 \frac{k}{6} a^{k-6} .
\end{equation}
(Recall $k=3(1-w)$.)
After substituting $a$ by $\phi$ and using (\ref{EQ_PHI_OF_A}) we find
\bdm
V(\phi) = \rho_0 \frac{k}{6} \left[ \sqrt{\frac{\kappa_+^2 +
      \kappa_-^2}{2\rho_0}} \sinh\left( \sqrt{\frac{3}{2 l (6-k)}} k
    |\phi| \right) \right]^{2\frac{k-6}{k}};
\edm
compare Fig.~\ref{fig:scalar_potential}. Potentials with
sinh-functions also occur frequently in FLRW models
\cite{Paulo,Mariam}. 

As the Wheeler-DeWitt equation will not
be analytically solvable for a general potential, we choose here $k=12$
($w=-3$, $l=-1$) as a particular example. This is, on the one hand, simply
solvable and reflects, on the other hand, the general case. We
approximate $V(\phi)$ and  
$|\phi(a)|$ for large $a$ and therefore large $|\phi|$; that is, we
investigate the limit when approaching the big rip. This gives 
\begin{equation}
\begin{aligned}
V(\phi) \sim \sqrt{\frac{\rho_0}{2}(\kappa_+^2 + \kappa_-^2)} ~
\ee^{6|\phi|} , \\ 
|\phi| \sim \frac{1}{6} \log\left[ \sqrt{\frac{8\rho_0}{\kappa_+^2
      + \kappa_-^2}} \right] + \alpha . 
\end{aligned}
\label{EQ_V_PHI_APPROX}
\end{equation}

Let us now turn to quantum cosmology. Note that the DeWitt metric 
has here signature $(-,-,+,+)$, since the kinetic term of the (phantom)
scalar field has the same sign as the one of the scale factor
(cf. Eq.~(30) in \cite{DKS}). The conformally covariant
Wheeler-DeWitt equation with the scalar potential in the limit
approaching the big rip reads 
\begin{equation}
\left[
\dif{}{\alpha}{2}
 -
\dif{}{\beta_+}{2}
 -
\dif{}{\beta_-}{2}
 + 
\dif{}{\phi}{2}
+V_0 \ee^{6(\alpha + |\phi|)}
\right]\Psi=0,
\label{eq:WDW_phantom}
\end{equation}
where $V_0 := \sqrt{2\rho_0 (\kappa_+^2 + \kappa_-^2)}$. After
intermediate steps in which one makes use of the variables
$u:=\alpha + |\phi|$ and $v:=\alpha -
|\phi|$, we solve the equation using a separation ansatz. Then the full
solution is 
\begin{equation}
\begin{aligned}
&\Psi \left(
\alpha,
\beta_+,
\beta_-,
\phi 
\right)
=
\\
&\sum\limits_{\sigma=\pm}
\int_{\mathbb{R}^3} \dd p_+ 
\dd p_- \dd p_3 \ 
\mathcal{A}_\sigma\left(p_+,p_-,p_3 \right)
\\
&
\quad\times\psi_{p_+,p_-,p_3}^\sigma \left(
\alpha,
\beta_+,
\beta_-,
\phi 
\right)
\end{aligned}
\end{equation} 
with mode functions
\begin{equation}
\begin{aligned}
& \psi^{\pm}_{p_+,p_-,p_3}
\left(\alpha,\beta_+,\beta_-,\phi \right)
=  \ee^{\ii p_+ \beta_+ + \ii p_- \beta_-} 
\\
&  \times
H_{\frac{p_3}{3}}^{(1,2)} \left( \sqrt{\frac{V_0}{18}} 
\ee^{3(\alpha + |\phi|)} \right)
\\
&  \times
\exp\left( \mp \ii \sqrt{\frac{p_+^2 + p_-^2}{2} + p_3^2}
  (\alpha - |\phi|) \right),
\label{EQ_BIGRIP_MODE}
\end{aligned} 
\end{equation}
where $H_\nu^{(1,2)}(z)$
are the Hankel functions. Note that the latter assume the WKB form for large
arguments, where for the van Vleck determinant $D
\propto \ee^{-3(\alpha+|\phi|)}$ holds. From (\ref{EQ_V_PHI_APPROX}) we see
that $|\phi| \propto \alpha$ for large $\alpha$. Thus the amplitude of
the wave function 
decreases and the wave function vanishes as we approach the big rip
singularity. As in the previous section, the DeWitt criterion
is fulfilled, and the singularity is avoided if this criterion is adopted.

Using the asymptotic WKB form of the Hankel functions, we get a WKB
solution for the complete mode wave function. The phase is 
\begin{equation}
\begin{aligned}
S = & \pm \left( \sqrt{\frac{V_0}{18}} \ee^{3(\alpha + |\phi|)} -
  \frac{\pi}{6} p_3\right.
  \\
  & \quad
  \left.- \sqrt{\frac{p_+^2 + p_-^2}{2} + p_3^2} (\alpha -
  |\phi|) \right) 
\\
& \quad + p_+ \beta_+ + p_- \beta_- .
\end{aligned}
\end{equation}
Using the principle of constructive interference \cite{Wheeler68,Kiefer88}, we get
\begin{equation}
\begin{aligned}
|\phi| &= \frac{\pi}{6} \sqrt{1 + \frac{p_+^2 + p_-^2}{2 p_3^2}} + \alpha, \\
\beta_\pm &= \pm \frac{p_\pm}{\sqrt{2(p_+^2 + p_-^2) + 4 p_3^2}}
(\alpha - |\phi|). 
\end{aligned}
\end{equation}
As expected, we find that $|\phi| \propto \alpha+{\rm const.}$ and
that $\beta_\pm$ 
become constant. Note that we have not recovered here 
(\ref{EQ_BETA_OF_A}), because we have used 
an approximated form of the potential and an asymptotic expression of the
wave function.  
\begin{figure}[!ht] \centering
  \begin{subfigure}[b]{0.48\textwidth}
    \centering
    %\vspace{ 0.5cm}
    \includegraphics[width=\textwidth]{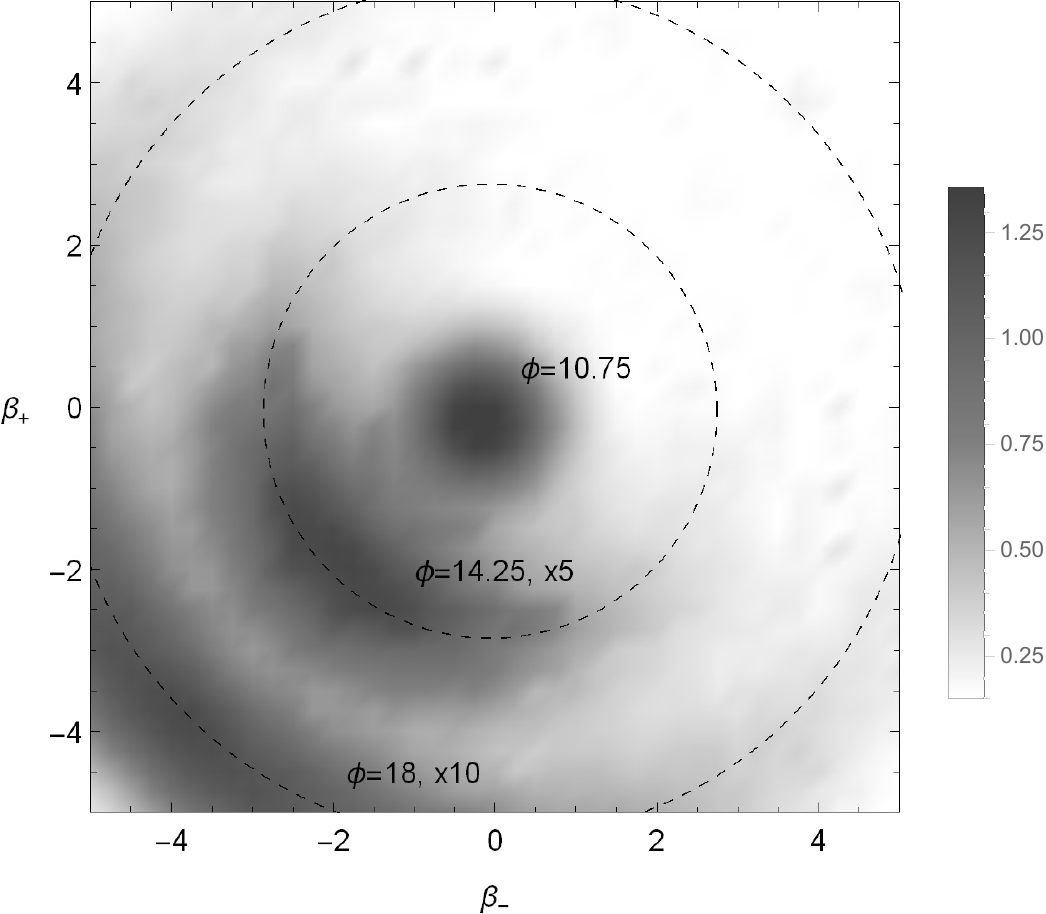}
    \caption{Density plot for $|\Psi|
      \ee^{\frac{3}{2}(\alpha+|\phi|)}$ showing the position of the
      wave packet for different values of $|\phi|$, each with a
      different scaling to visualize the decaying wave in a single
      graphic.}  
    \label{fig:bigrip_cut}
  \end{subfigure}%
  \vspace{0.5cm}
  \begin{subfigure}[b]{0.45\textwidth}
    \centering
    \includegraphics[width=\textwidth]{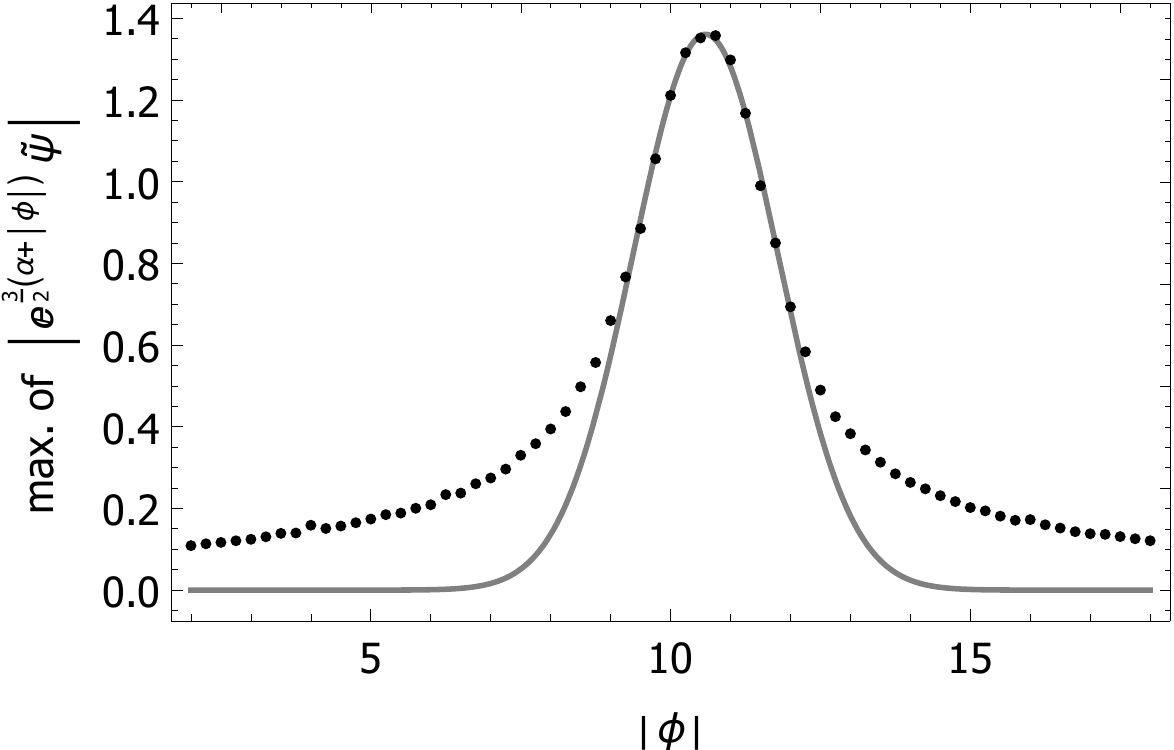}
    \caption{Maxima for $|\Psi|
      \ee^{\frac{3}{2}(\alpha+|\phi|)}$ for different $|\phi|$
      (dotted) with a Gaussian fit to the peak (solid).}  
    \label{fig:bigrip_decay}
  \end{subfigure}
  \caption{Numerical results for the wave packet $|\Psi|
    \ee^{\frac{3}{2}(\alpha+|\phi|)}$ using the asymptotic WKB form of
    (\ref{EQ_BIGRIP_MODE}) and $\alpha=10$,
    $\mathcal{V}_0=\bar{p}_+=\bar{p}_-=\bar{p}_3=1$, $\Delta p=1.5$.} 
  \label{fig:bigrip_all}
\end{figure}

From now on we set $\mathcal{A}_- = 0 $ and choose a Gaussian weighting function for $\mathcal{A}_+$, that is,
\begin{equation}
\begin{aligned}
\mathcal{A}_+\left(p_+,p_-,p_3 \right) =
 \frac{\ee^{ -\frac{(p_+ - \bar{p}_+)^2 + (p_- - \bar{p}_-)^2 + (p_3 -
       \bar{p}_3)^2}{2\Delta p^2} }}{(\sqrt{2\pi} \Delta p)^3},  
\end{aligned}
\end{equation}
where $\bar{p}_{\pm,3}$ are non-zero mean values, and $\Delta p$
denotes the
width. To perform the integrals analytically, we assume that the
momenta are sharply peaked around their mean values such that we can
linearize the dispersion relation 
\begin{equation}
\begin{aligned}
& \sqrt{\frac{p_+^2}{2} + \frac{p_-^2}{2} + p_3^2} \approx \frac{1}{2}
\frac{\bar{p}_+}{\bar{p}} p_+ + \frac{1}{2} \frac{\bar{p}_-}{\bar{p}}
p_- + \frac{\bar{p}_3}{\bar{p}} p_3,  
\\
& \bar{p} := \sqrt{\frac{\bar{p}_+^2}{2} + \frac{\bar{p}_-^2}{2} + \bar{p}_3^2},
\end{aligned}
\label{eq:linearization}
\end{equation}
and use the WKB limit of the mode functions. Using the formula for
Gaussian integrals we end up with

\begin{widetext}
\begin{equation}
\begin{aligned}
\Psi(\alpha,\beta_+,\beta_-,\phi) &= \exp\left[
  -\frac{\Delta p^2}{2} \left[ \left(\beta_{+} -
      \frac{\bar{p}_{+}}{2\bar{p}} (\alpha-|\phi|) \right)^2 +
    \left(\beta_{-} - \frac{\bar{p}_{-}}{2\bar{p}} (\alpha-|\phi|)
    \right)^2 + \left(\frac{\pi}{6} + \frac{\bar{p}_3}{\bar{p}}
      (\alpha-|\phi|) \right)^2 \right] \right] \\ 
&\times \exp\left[ \ii \left( \bar{p}_{+} \left( \beta_{+} -
      \frac{\bar{p}_{+}}{2\bar{p}}(\alpha-|\phi|)\right) + \bar{p}_{-}
    \left( \beta_{-} - \frac{\bar{p}_{-}}{2\bar{p}}(\alpha-|\phi|)
    \right) - \bar{p}_3 \left( \frac{\pi}{6} +
      \frac{\bar{p}_3}{\bar{p}}(\alpha-|\phi|) \right) \right)
\right] \\ 
&\times \exp\left[\ii \sqrt{\frac{\mathcal{V}_0}{18}}
  e^{3(\alpha+|\phi|)} - \ii \frac{\pi}{4} \right] 
  \times \left( \frac{72}{\mathcal{V}_0 \pi^2} \right)^\frac{1}{4} \ee^{-\frac{3}{2}(\alpha+|\phi|)}. 
\label{EQ_BIGRIP_WAVEPACKET_LINEAR}
\end{aligned}
\end{equation}
\end{widetext}

As before, the wave packet decreases due to the presence of the van
Vleck determinant. It is peaked around the
classical trajectory without dispersion, with a slight modification
coming from the van Vleck determinant.
A similar behavior was found in \cite{Kiefer88} for a massless scalar
field in a FLRW universe.

The linearization \eqref{eq:linearization} breaks down when the absolute value of
$\alpha-|\phi|$ becomes large. We then perform
a numerical integration of the full wave packet\footnote{The wave
packet is rescaled by the inverse of the van~Vleck factor.}
$|\Psi|\ee^{\frac{3}{2}(\alpha+|\phi|)}$. Fig.~\ref{fig:bigrip_cut}
shows the results for 
$\alpha=10$ and different values of $|\phi|$. For $|\phi|\approx
10.75$, the wave packet has a global maximum which corresponds to the
analytical result (\ref{EQ_BIGRIP_WAVEPACKET_LINEAR}). For increasing
$|\phi|$, the wave packet assumes an annular shape and propagates
outwards with 
decreasing amplitude. The wave is peaked in the direction of negative
$\beta_\pm$. For decreasing $|\phi|$, one has a similar
behavior with the maximum moving into the opposite direction. Note
that this annular waves also appear for the corresponding wave packet
of the Kasner solution. The dispersion takes place due to the
additional degrees of freedom introduced by the anisotropy.  

In Fig.~\ref{fig:bigrip_decay} we display the maxima of the wave packet 
for different $|\phi|$ together with a Gaussian fit to the peak
region.  One can see that close to the peak the wave packet decreases
like a Gaussian, 
but decays much weaker (not even exponentially)
further away. 
The amplitude of the full wave packet $|\Psi|$ including the van~Vleck
determinant will increase for decreasing
$|\phi|$ such that the peak along the classical trajectory will be at
best a
local maximum. This might be interpretable as a transition
from a semiclassical into a full quantum regime.

The Klein-Gordon flux does not vanish.
Similar to the case considered in the previous section, the exponential function  from
the van Vleck determinant cancels, yielding 
\begin{widetext}
\begin{equation}
\begin{aligned}
& J^{\alpha,\phi} \propto \left| \int_{\mathbb{R}^3} \dd p_+ \dd p_-
  \dd p_3 \mathcal{A}_+(p_+, p_-, p_3) ~ \exp \left[ \ii \left( p_+
      \beta_+ + p_- \beta_- - \sqrt{\frac{p_+^2 + p_-^2}{2} + p_3^2}
      (\alpha + |\phi|) - \frac{\pi}{6} p_3 \right) \right] \right|^2
+ \mathcal{O}\left( \ee^{-3(\alpha+|\phi|)} \right) \\ 
& J^{\beta_\pm} = \mathcal{O}\left( \ee^{-3(\alpha+|\phi|)}
\right), 
\end{aligned}
\end{equation}
\end{widetext}
where $J^I=\frac{1}{2\ii}\mathcal{G}^{IJ}\left( \Psi^*
    \partial_I \Psi- \Psi \partial_I \Psi^* \right)$, and
  $J^{\alpha,\phi}$ means that both $J^{\alpha}$ and $J^{\phi}$ 
 are proportional to the expression on the right-hand
 side. Numerically we find for it a 
similar structure as for
$|\Psi|\ee^{\frac{3}{2}(\alpha+|\phi|)}$, that is, the Klein-Gordon flux is peaked over the classical trajectory. According to criterion~2,
the big rip is not avoided.
Similar to the case of the last section, the singularity is thus only
avoided by the DeWitt criterion.
This explicit calculation shows, moreover, that 
the wave packet is not 
peaked all along the classical trajectory if one considers
$|\Psi|$, whereas it is peaked
with respect to the Klein-Gordon flux.

\section{Conclusion}

In this paper, we have extended previous investigations of singularity
avoidance from isotropic to anisotropic models. We have, in
particular, adapted the avoidance criterion to the covariant structure
of minisuperspace, which becomes relevant for dimensions higher than
two. We have found that the DeWitt criterion can, in general, be
fulfilled, but not so the vanishing of the Klein-Gordon current. For
the reasons mentioned, however, we attribute more relevance to the
DeWitt criterion.

The flux criterion was recently applied in
  \cite{DS19} to investigating the fate of (big bang and big crunch)
  singularities in FLRW models with Brown-Kucha\v{r} dust. Singularity
  avoidance was then found for a certain class of factor orderings. 

In our paper, singularity avoidance was found by studying
  properties of the quantum cosmological Wheeler-DeWitt equation. 
  The structure of this differential equation was used to disclose
  the fate of both the big bang and the big rip singularities. The
  tacit assumption in this is that information about the fates of
  different types 
  of singularities can be obtained from one and the same differential
  equation (in the same way as information about different types of
  classical singularities can be obtained from the same
  Friedmann-Lema\^{\i}tre equations). A full mathematical treatment
  should address the properties of the Wheeler-DeWitt equation and its
  boundary conditions in much more detail, pointing out structural
  differences between the singularities, but this is beyond the
  scope of this paper.

While the general criteria were formulated for general  Bianchi class A
models, detailed investigations were made for the Bianchi~I model with
and without matter. Bianchi~I models admit the prototype of an
asymptotically velocity term dominated (AVTD) model. Our results of
singularity avoidance should thus be representative for such a kind of
singularity with a sufficiently large number of degrees of freedom. 
Other Bianchi models such as Bianchi~VIII and Bianchi~IX exhibit an
oscillatory behavior when approaching the singularity. 
The singularity can, however, become AVTD
if, for example, a scalar field is added \cite{Berger}.

Because Bianchi~IX models are generally considered as reflecting
the generic behavior towards a spacelike singularity, future
investigations of singularity avoidance should attempt to address
these models in as much detail as possible. For this, it would be
desirable to have mathematical theorems available such as those
discussed here for the Kasner solution. In \cite{Ringstroem_Book} one
finds an existence and uniqueness theorem (Theorem 8.6 there), but for
the Bianchi~IX potential no decay rate estimates seem to exist.
The Wheeler-DeWitt equation for the vacuum Bianchi~IX (mixmaster)
model was solved numerically in \cite{Koehn} by using the `hard wall
approximation'. The results found in this analysis strongly indicate
the decay of wave packet amplitudes.  

It is generally believed that the approach to a spacelike singularity
at the level of the full Einstein equations can be described by the
Belinsky-Khalatnikov-Lifshits (BKL) scenario; see, for example,
\cite{Berger,KKP18} and references therein. This corresponds to a 
decoupling of spatial points, in which every spatial point exhibits
the dynamics of a separate Bianchi IX model. The eventual goal will be to
present  a quantum-gravitational 
analysis of this situation, from which one should be able to draw
general conclusions about singularity avoidance. An investigation in
the framework of affine quantization was made recently in
\cite{GPP19}. Attempts in this
direction using the Wheeler-DeWitt equation will be the subject of future
investigations.

\section*{Acknowledgments}

We are grateful to Anne Franzen and Tim Schmitz for helpful discussions.

%%%%%%%%%%%%%%%%%%%%%%%%%%%%%%%%%%%%%%%%%%%%%%%%%%%%%%%%%%%%%%%%%%

%%%%%%%%%%%%%%%%%%%%%%%%%%%%%%%%%%%%%%%%%%%%%%%%%%%%%%%%%%%%%%%%%

\end{document}